\documentclass[preprint,aps,amsmath,amssymb, floatfix]{revtex4}
\usepackage{bm}% bold math
\usepackage{amssymb}
\usepackage{amsmath}
\def\be{\begin{equation}}
\def\ee{\end{equation}}
\def\beq{\begin{eqnarray}}
\def\eeq{\end{eqnarray}}

\begin{document} %\openup 8pt %\preprint{smw-let-01-09}
\title{Emission Origin for the Wave of Quanta}
\author{Sanjay M Wagh}
\affiliation{Astrophysics \& Cosmology Research Unit, University of KwaZulu-Natal, Private Bag X54001
\\ Durban 4000, South Africa\footnote{Telephone: +27 (0)31 260 3000 (pp) \\ Fax: +27 (0)31 260
2632 \\  Electronic address: waghsm.ngp@gmail.com;
waghs@ukzn.ac.za}}

%\altaffiliation{Permanent address: Central India Research Institute, Post Box 606,
%Laxminagar, Nagpur 440 022, India}  %\vspace{1in}}

\begin{abstract}
We argue that certain assumptions about the process of the emission
of the quanta by their (oscillating) emitter provide for their
changing (oscillatory) flux at any location. This mechanism
underlying (such) wave phenomena is not based, both, on the
newtonian notion of force and the field concept (of Faraday,
Maxwell, Lorentz and Einstein). When applied to the case of thermal
radiation, this emission origin for the wave of quanta is shown here
to be consistent with the laws of the black body radiation. We
conclude therefore also that a conceptual framework, which is not
rooted in the notion of force and in the field concept, may provide
a deterministic basis underlying the probabilistic methods of the
quantum theory.
 \bigskip



%\noindent{{\bf PACS:} 01.55+b (General Physics) - 03.65.Ta (Foundations of QM) - 05.30-d (Quantum Statistical %Mechanics)}

\end{abstract}
\date{June 24, 2009}
\maketitle
%%%%%%%%%%%%%%%%%%%%%%%%%%%%%%%%%%%%%%%%%%%%%%%%%%%%%%%%%%%%%%%%%
\section{Introduction} \label{intro}
In contrast to Maxwell's theory of the monochromatic radiation as
being a transverse and propagating wave of electromagnetic nature,
Planck assumed \cite{planck} radiation as quantal during
interactions with matter. In 1900, he also assumed that the quantal
energy $\epsilon$ is related to the frequency $\nu$ of (the
electromagnetic wave of) monochromatic radiation as $\epsilon = h
\nu$. Planck's law for the spectral energy density of radiation in
thermal equilibrium agreed with experiments, then. Maxwell's theory
could explain only the Rayleigh-Jeans part of Planck's spectral
energy density function, however.

In extending Planck's reasoning, Einstein assumed that monochromatic
radiation behaves like a discrete medium consisting of quanta
carrying energy $\epsilon = h \nu$ also in transits. Einstein thence
explained \cite{ein-photo} the photoelectric effect in 1905. He had
further envisaged \cite[p.404]{pais-a}, in 1909, that {\em the next
phase in the development of theoretical physics will bring us a
theory of light that can be interpreted as a kind of fusion of the
wave and the emission theory $\cdots$ [The] wave structure and [the]
quantum structure are not to be considered as mutually incompatible
$\cdots$}

Theory of spectra was proposed by Bohr \cite{bohr} in 1913. Then, in 1916, Einstein %, in 1916,
had considered \cite{return01} \cite{return02} \cite{return03} an
atomic gas in thermal equilibrium with radiation. Following Bohr, he
assumed that atoms of gas make transitions from one to another of
their energy levels by emitting or absorbing the quanta of
monochromatic radiation. In addition to processes of induced
emission and absorption of quanta, Einstein was led to postulate the
process also of their spontaneous emission so that Planck's law is
obtained for the spectral energy density of radiation in this
situation.

In his works of 1916, Einstein's treatment of the spontaneous
emission of quanta of radiation was statistical. He then pointed to
the similarity of this statistical theory with Rutherford's theory
for the radioactivity \cite{radioactivity}, which also involves
spontaneous emission of particles, or quanta of matter, by a single
atom.

Many physicists, including Einstein, expected a deterministic
framework to be the basis of statistical theories. However, no one
could show how the statistical treatment could be supported with a
deterministic framework for the process of the spontaneous emission
of quanta, either of radiation or of matter. \newpage

Consequently, Einstein remained unsatisfied with the statistical
nature of his theory of spontaneous emission of quanta of light. In
1951, he expressed this dissatisfaction by saying that: {\em All
these fifty years of pondering have not brought me any closer to
answering the question, What are light quanta?}

Today, the word ``photon'' (coined, firstly, in \cite{lewis}) is
synonymous with a quantum of light of vanishing rest-mass, momentum
$h\nu/c$, and unit intrinsic spin.

Now, Maxwell's theory indicates that the quanta of radiation be
inertia-less.  Then, just like their inertia is, their momentum must
be vanishing, if defined as a product of inertia and velocity. But,
Compton-like effects \cite{compton01} \cite{compton02} can be
explained by assuming that the inertia-less quanta of radiation
possess non-vanishing momentum $p=h\nu/c$. Such considerations
appear to be mutually contradictory, however.

Notably, the non-vanishing of the photon momentum $p=h\nu/c$ appears
to arise out of Planck's relation $\epsilon = h \nu$ alone. Grasping
the nature of Planck's relation should thus be pivotal to our
understanding of the quantum phenomena.

With this aim, we assume certain aspects of the emission of quanta
by their emitter. Propagating spherical fronts of quanta emitted at
different instances are then centered about different locations of
their (oscillating) emitter. This leads to the changing (wavy)
nature of the quantal flux at any location, then. Emission aspects
thus constitute the {\em emission origin\/} for the changing (wavy)
nature of the flux of quanta.

From the perspective of this emission origin for the wavy flux of
quanta, we may then focus on the situation of radiation contained
within a cavity and in thermal equilibrium with the walls of that
cavity. In this situation of the thermal equilibrium of radiation
with a cavity, or that of the black body radiation, we must obtain
Planck's law for the spectral energy density of radiation. In other
words, if the wave phenomena of radiation arise indeed due to
aspects of emission of quanta, then such an origin must be
consistent with Planck's law.

If the above is indeed the case, then we could expect Planck's
relation $\epsilon=h\nu$ to arise from these considerations. This is
the subject of the present study, which attempts to comprehend
Planck's $\epsilon=h\nu$ relation in this manner.

\section{Statistical Considerations} \label{stats}
In what follows, we recall the statistical basis of Planck's law for
the spectral energy density of the black body radiation. (See
\cite{stat-mech} for concerned details.) However, we will not assume
Planck's relation $\epsilon=h \nu$ to begin with.

For us, noteworthy here is S N Bose's derivation \cite{bose}  of
Planck's formula for the spectral energy density of the black body
radiation. Bose considered a gas of (inertia-less) particles of
(non-kinetic) energy $\epsilon=h\nu$. He then distributed these
particles in energy boxes, and computed their most probable
distribution.

(The term ``box'' will refer to the momentum part of the phase space
of the system, and will not be used to represent its spatial
dimensions.)

Bose did not explain how a corpuscular or quantal property
$\epsilon$ can be so related to a wave property $\nu$ as in the
relation $\epsilon=h\nu$. Nevertheless, assumptions underlying his
derivation of Planck's formula are of our interest here.

Bose considered {\em indistinguishable\/} objects, each with an {\em
individuality\/} of its own. (For an interesting discussion related
to these notions, see \cite{cheng}.) Number $N$ of these objects are
to be distributed in number $m$ of boxes, with the boxes being
labeled by energy as $E_i=(i-1)\,\epsilon$. Here, $\epsilon$ is a
constant; and the index $i$ takes integer values as $i=1, 2, \cdots,
m$. Then, a specific state of this system of $N$ objects has $n_i$
objects in the $i$-th box with $\sum_{i=1}^m n_i=N$. The number $N$
is not necessarily a constant. Both $N$ and $m$ are assumed to be
very large integers.

Bose had also assumed additional conditions leading to the
statistical  independence of objects while getting distributed in
energy boxes. He assumed also that objects do not prefer any
particular box. We also know that, for the case of radiation, it is
necessary to allow a single box to be populated by more than one
object.

Then, as can be easily shown using the canonical distribution, in
the most probable state of the system, the mean occupation number of
the i-th box is \[ \overline{n}_i =  1/\left( e^{\epsilon /\Theta}-1
\right) \] Then, in the most probable state of the system, the mean
energy of the i-th box is \[ \overline{E}_i = \epsilon/\left(
e^{\epsilon /\Theta}-1 \right) \] Here $\Theta\equiv kT$ is called
as the {\em modulus of the distribution}, with $T$ as the
temperature of the system and $k$ as Boltzmann's constant.

As can also be easily shown, when in the most probable state, the
system displays the mean square fluctuations in number given by \[
\overline{n_i^2} = \overline{n}_i\,\left[ \, 1 + \overline{n}_i
\,\right] \] When the system is in the most probable state, the mean
square fluctuations in energy are then given by \[ \overline{E_i^2}
= \overline{E}_i\, \left[ \,\epsilon + \overline{E}_i \, \right] \]
Now, these fluctuations, when the system is in the most probable
state, are {\em entirely statistical results}. These are to be
``interpreted'' in terms of the physical motions of objects (whether
the inertia of these objects is vanishing or not).

The most probable state of the system is now taken to be its
equilibrium state. For the physical problem of the radiation
enclosed in a cavity, the most probable state is then the state of
thermal equilibrium of radiation.

Consider the first term of the equilibrium fluctuations in energy
(or in the number of quanta) of radiation. Quanta have been assumed
to possess an individuality of their own. Then, we expect that the
fluctuations in energy, as in the first term, arise due to
collisions of quanta. Thus, this term is known as the {\em particle
term}.

For thermal radiation in a cavity, the second term of these
equilibrium fluctuations can be interpreted as being due to the wave
character of radiation as per Maxwell's theory. It is therefore
called as the {\em wave term}.

Wavy motions of objects may lead to the second term of equilibrium
fluctuations, also. This would require appropriate (Hooke's law)
forces acting on objects. But, no physical mechanism producing such
forces exists within this situation.

Interactions of quanta of radiation with the ``matter of the walls''
of the cavity could be considered as the subject of {\em the
emission theory\/} that Einstein referred to in 1909. But, {\em
forces causing wavy motions of quanta all over the space within the
cavity cannot arise even from such interactions}.

In summary, objects of Bose's considerations have an {\em
individuality\/} of their own, {\em ie}, they are particulate. The
second term of equilibrium fluctuations cannot then be due to their
being waves, or due to their wavy motions. Then, what is the origin
of these fluctuations? In what follows, we address this issue for
radiation, all whose quanta are assumed to move with the same speed.

\section{Emission and the Formation of Wave of Quanta}  \label{emits}

Now, we show that certain assumptions about the nature of the
emission of quanta explain the second term of their equilibrium
fluctuations. Wavy changes in the flux of quanta arise out of these
assumptions about the nature of their emission. However, forces are
not invoked to act on these quanta. We expect all the wave
characteristics of quantal propagation to emerge from these emission
aspects.

To this end, let the emission of quanta not occur unless the kinetic
energy of their emitter changes. Change in the kinetic energy of the
emitter is then assumed to be distributed among the emitted quanta
as their (non-kinetic) energy.

Now, as per Maxwell's theory, an oscillating charge produces an
electromagnetic wave propagating in all directions away from that
charge. This wave emission has the property that any wavefront is
spherically symmetric about the instantaneous position of the charge
at the instant of its emission. To be in conformity with this and,
thus, with many experiments, we assume that the propagating front
formed by the simultaneously emitted quanta is spherically symmetric
about the instantaneous position of the source at their emission.
(Notably, this assumption is also not inconsistent with the laws of
radioactivity, if the quanta are those of matter.)

In what follows, we use these assumed peculiarities of the emission
of quanta to show that Planck's law obtains for their spectral
energy density, when in thermal equilibrium with the walls of the
cavity containing them.

In the above context of the emission of light-quanta by their
emitter, the following questions then arise. How many quanta get
emitted at an instant? How does the kinetic energy of the emitter
get distributed among quanta? What are the laws underlying the
process of the emission of quanta? Issues arising out of such
questions are related to aspects of Maxwell's theory dealing with
the power radiated by an accelerated charge. These will be the
subject of a later communication, however.

We are not requiring, at this stage here, the details of the laws of
emission of quanta as well as those of how the accelerated motion of
the emitter is arising. Presently, we only need to know that the
emitters of quanta undergo oscillatory motions at the cavity walls.
This is so for the following reasons.

We assumed that quanta are emitted only when the kinetic energy of
their emitter changes. For the change $\triangle E$ in the kinetic
energy of the emitter, let it emit number $n$ of light-quanta at
that instant. (When the energy $\triangle E$ is distributed equally
among the emitted $n$ quanta, each quantum has energy
$\epsilon=\triangle E/n$.)

After one emission of quanta, the second emission takes place when
the kinetic energy of the emitter changes next. But, this second
spherical front of quanta is now centered at a different location of
their emitter. Then, passing any specified or given location, the
flux of quanta due to the first emission front would be different
than that due to the second emission front.

The changing flux of quanta at a given location can now arise due to
the following two reasons. Firstly, the number of quanta emitted at
the instances of the emission of two different spherical fronts
could be different. Secondly, for the given location of the unit
area, its angle subtended at the location of the emitter would be
different, spherical fronts being centered at different locations.

Clearly, for oscillatory changes in the location of the emitter,
oscillatory changes would be seen in the flux of quanta at any
location as well. Then, the frequency of oscillations of the number
of quanta at any arbitrary location is also the frequency of
oscillations of their emitter.

Assumed peculiarities of the emission of light-quanta are
responsible for the wavy behavior of their flux passing any
location, therefore.

No light-quantum is undergoing wavy motion. We do not therefore
require (Hooke's type) forces causing the wavy motions of quanta.
But, the wavy behavior of the flux of quanta results. It is due to
the oscillatory motion of their emitter. Such wave aspects of
quantal propagation thus originate in their emission aspects.
\newpage

Now, for the case of radiation in thermal equilibrium with the walls
of a cavity, an emitter of quanta is necessarily required to be a
linear oscillator. If this were not the situation, then we would
expect the density of radiation within the cavity to be varying with
time. This contradicts our assumption of the thermal equilibrium of
radiation with the walls of the cavity containing it.

Planck had assumed linear oscillators for matter at the walls of the
cavity. We too have {\em linear oscillators\/} at the walls of the
cavity in thermal equilibrium with radiation, then. Planck's law for
the cavity radiation should thus be obtainable within the premise of
the present considerations.

Now, for the radiation in thermal equilibrium within a cavity, let
us note that the wave modes of radiation would be standing modes.
Let us also assume two polarization states for these wave modes. (We
offer no explanation for this assumption. But, it should relate to
results established in \cite{wigner}.)

Consider now a cubical cavity of each side of length $\ell$. Assume
the wavelength $\lambda$ of the standing wave mode to be smaller
than the cavity length-scale $\ell$. Under equilibrium, the number
$f(\nu)\delta\nu$ of standing wave modes within the cavity and
within the frequency range $\nu$ to $\nu+\delta \nu$ is, then, \[
f(\nu)\delta\nu = \frac{4\ell^3\nu^2}{c^3}\delta\nu \] with $c$
being the wave speed. Then, the number $\delta n_{\nu}$ of
light-quanta {\em involved in the wavy flux\/} and within the
frequency range $\nu$ to $\nu+\delta \nu$ is, simply, given by \[
\delta n_{\nu} = \;\overline{n}_i\;f(\nu) \delta\nu \] Then, the
energy of radiation within the same frequency range is
\[ \delta E_{\nu} = \epsilon \;\delta n_{\nu} \] The spectral energy
density of the cavity radiation is therefore obtained as  \[
\frac{\delta E_{\nu}} {\ell^3}  = \frac{(4\epsilon \nu^2/c^3)
\delta\nu} { e^{\epsilon /kT} - 1} \]

For $\epsilon = h \nu$, this expression now yields Planck's law for
the black body radiation. Planck's law is consistent with the
corresponding equilibrium fluctuations. Therefore, our
considerations are consistent with those fluctuations, too.

{\em Emission origin for the wavy nature of the number of quanta at
any location in a cavity in thermal equilibrium with radiation is
then consistent with the laws of the black body radiation}. The
second term of the statistical fluctuations of the most probable
state of cavity radiation arises due to the wavy nature of the
``number'' of quanta at a cavity location. But, the wavy flux is due
to aspects of the emission of quanta at the walls of a cavity they
are in thermal equilibrium with.

We needed to resolve the question of the origin of the wavy
fluctuations in the number of quanta within a cavity, without
invoking ``forces'' to act on them. Aspects of the emission of
quanta provide such a resolution of this question.

\section{Concluding Remarks} \label{conclude}

In summary, we assumed that quanta move with constant velocity till
interactions. For radiation, speed is the same for all the quanta.
Emitter of quanta was assumed to undergo a change $\triangle E$ in
its kinetic energy for the quantal emission. The energy $\triangle
E$ was assumed to be distributed (as non-kinetic energy) among the
quanta emitted at the same instant. Emission of quanta was assumed
to be spherically symmetric about the position of the emitter at the
instant of emission.

Then, the flux of quanta at any location changes as per the motion
of the emitter. If the motion of the emitter were oscillatory with
frequency $\nu$, then the flux of quanta at any location would also
be oscillatory with the same frequency $\nu$.

We showed here that this ``emission origin'' for the wavy behavior
of the flux of light-quanta is consistent with the laws of the black
body radiation.

However, the (non-kinetic) energy $\epsilon$ of any emitted
light-quantum is {\em not related\/} here to the frequency $\nu$ of
the material oscillator of the walls of a cavity that the radiation
is assumed to be in thermal equilibrium with. Planck's relation
$\epsilon = h \nu$ appears, therefore, to be an ad-hoc assumption
relating a corpuscular property $\epsilon$ to a wave property $\nu$.
Consequently, Planck's constant $h$ too has an ad-hoc status, here.

Why is Planck's postulate $\epsilon=h\nu$ consistent with
experimental results then? To comprehend this, we recall here our
assumption that no quanta are emitted unless the kinetic energy of
the emitter changes. Then, the quantal energy $\epsilon$ is
vanishing when the frequency $\nu$ of the oscillator is vanishing.
As entirely mathematical relationship, without any physical basis,
we may postulate their proportionality. This we understand as a
reason for $\epsilon=h\nu$ being consistent with the observed laws
of the black body radiation. On the basis of present considerations,
we therefore ``comprehend'' Planck's this relation as being an
ad-hoc postulate.

(We would like to note here that the ad-hoc status of Planck's
celebrated constant is consistent with Einstein's views about %regarding
the fundamental role of constants of Nature in a theory \cite[p.
74]{ein-quote}: {\em In a sensible theory there are no
[dimensionless] numbers whose values are determinable only
empirically. I can, of course, not prove that $\cdots$ dimensionless
constants in the laws of nature, which from a purely logical point
of view can just as well have other values, should not exist.})

Now, our problem of the radiation existing in thermal equilibrium
with the walls of a cavity has two aspects, namely, the
emission/absorption of radiation at the walls and the propagation of
radiation inside the cavity. We established here that the wavy flux
of propagating quanta of radiation originates in aspects of their
emission by material oscillators at the walls of the cavity
containing them.

If we focus on only one of these two aspects and neglect the other,
then we would be {\em dividing this problem into two parts - the
wave part and the emission part}. We may then attempt to describe
them separately of each other. This would be the {\em divide and
conquer strategy of problem-solving}. Within this strategy, we need
to establish the mutual consistency of the separate solutions of the
two parts, however.

Any emitter of quanta in thermal equilibrium with radiation has an
oscillatory motion. It is due to continual emission and absorption
of the quanta of radiation. Under the ``divide and conquer
strategy'' of problem-solving, such motion could be described as
being due to a force acting on it. Such a force, the Lorentz force,
causing the motion of a radiating charge is the basis of Maxwell's
electrodynamics.

Under the same ``divide and conquer strategy'' of problem-solving,
the wave part can now be described in terms of a wave equation.
Then, a wave equation could be based on an appropriate (Hooke's
type) force, or on an appropriate substratum (field) providing a
medium or basis for the propagation of waves. \newpage

A physical mechanism causing wavy motions of the quanta of radiation
does not exist within the cavity situation. A wave equation could
then be based on the concept of a field. Maxwell's theory provided
such a wave equation.

However, such a framework of concepts was unable to explain the laws
of the black body radiation, the photoelectric effect, $\cdots$
Planck's radical and revolutionary hypothesis of the quantal nature
of radiation turned out to be necessary for explaining these
phenomena. Faraday, Maxwell and Lorentz's framework of concepts was
then replaced by the probabilistic one of the quantum theory.

Schr\"{o}dinger's equation now describes \cite{qm} the mechanics of
a material body, of non-vanishing inertia, using the concept of a
potential (arising out of that of force acting on that body). It
then provides for the motion of the emitter of the quanta of
radiation. Bohr's theory of spectra arises out of these
considerations.

Path to the description of the quanta of radiation in transit was,
however, long. It required a mathematical formalism providing for
their number, that is to say, a formalism consistent with their
quantum nature. The second quantization formalism of the quantum
theory is such a method \cite{dirac} to obtain the number of quanta
within a given wave mode. It is applicable to quanta of matter and
of radiation, both.

Quantum theory considers then only the wave aspect of the problem at
hand. We then ``comprehend'' the formalism of the quantum theory as
arising due to the ``divide and conquer strategy'' of
problem-solving. Based on only the wave part, the quantum theory
provides the probabilistic description for the quanta.

Clearly, the formalism of the quantum theory then completely misses
crucial aspects of the emission of quanta. Needless to say then, but
important also to point out here, the description of the physical
world provided by the quantum theory is, therefore, an incomplete
one in this respect.

In this context, we then note that of pivotal importance to
Einstein's thinking was Boltzmann and Gibbs's statistical theory of
many particle systems, which is incomplete in relation to Newtonian
mechanics. He considered the quantum theory to be similarly an
incomplete description of physical systems, as only a statistical
theory that is inherently incapable of describing the dynamics of a
single object of any many-body system. He therefore expected some
deterministic framework to be the basis underlying the methods of
the quantum theory. Then, from Einstein's aforementioned point of
view, Heisenberg's indeterminacy relations hold only statistically.

We emphasize that we have not used the concepts of force and field
to explain the wave of quanta. Notably, in 1954, Einstein \cite[p.
467]{pais-a} considered {\it it quite possible that physics cannot
be based on the field concept, i.e., on continuous structures.}

%Then, with remarkable openness for ideas, he had said further: {\it [In that case], nothing remains of my entire %castle in the air, gravitation theory included, [and of] the rest of modern physics.}

Here, continuous structures are to represent the ``objects''
themselves, and not the probability of their being at some space
location at some instant of time. This is the ``field'' concept of
Faraday, Maxwell, Lorentz and Einstein.

The present results then imply that Einstein's aforementioned
opinion may have merit, because we have not used Einstein's concept
of field to represent quanta and to also explain their wave. {\em It
should be noted that we have not even specified here which
mathematical structure represents objects of our statistical
considerations.}

In this context, spontaneous emission of quanta (of radiation or
not) by one atom may have to be dealt statistically, but the
computation of its probability can then have basis in deterministic
ideas not based on the concepts of force and field, both.

{\em Then, we also conclude here that a deterministic basis, which
is rooted neither in the newtonian notion of force nor in the field
concept, may underly the probabilistic methods of the quantum
theory}. This possibility was first explored by Hertz \cite{schlipp}
with an intention to free Newtonian mechanics from the notion of the
potential energy, which he considered as unsatisfactory a concept.

\begin{acknowledgements} I am grateful to many friends for discussions and
encouragement over a long period of time. Recently, I am grateful,
in particular, to Dan Krige, and to John Hey who suggested a number
of improvements and drew my attention to relevant literature.
\end{acknowledgements}

%\newpage


\begin{thebibliography}{99}

\bibitem{planck} Planck M.: Deutsch. Phys. Ges. 2, 202 (1900)

\bibitem{ein-photo} Einstein A.: Ann. der Physik 17, 132 (1905)

\bibitem{pais-a} Pais A.: Subtle is the Lord ... The Science and the Life of Albert Einstein. Clarendon Press, Oxford (1982)

\bibitem{bohr} Bohr N.: Phil. Mag. 26, 1 (1913)

\bibitem{return01} Einstein A.: Verh. Deutsch. Phys. Ges. 18, 318 (1916)

\bibitem{return02} Einstein A.: Mitt. Phys. Ges. Z\"{u}rich. 16, 47 (1916)

\bibitem{return03} Einstein A.: Zur Quantentheorie der Strahlung. Phys. Zeitschr. 18, 121 (1917)

\bibitem{radioactivity} Rutherford E. and Soddy F.: Phil. Mag. 4, 370, 569 (1902)

\bibitem{lewis} Lewis G. N.: The Conservation of Photons. Nature. 118, 874-875 (1926)

\bibitem{compton01} Compton A. H.: Phys. Rev. 21, 483 (1923)

\bibitem{compton02} Stuewer R. H.: The Compton Effect, Science History, New York (1975)

\bibitem{stat-mech} Lindsay R. B.: Introduction to Physical Statistics. Dover, New York (1941)

\bibitem{bose} Bose S. N.: Plancks Gesetz und Lichtquantenhypothese. Z. Physik. 26, 178-181 (1924)

\bibitem{cheng} Cheng Chi-Ho: Thermodynamics of the System of Distinguishable Particles. Available via
arxiv.org/physics.chem-ph/0903.4748

\bibitem{wigner} Wigner E. P.: Ann. Math. 40, 149 (1939)

\bibitem{ein-quote} Rosenthal-Schneider I.: Reality and Scientific Truth. Wayne State University Press (1980)

\bibitem{qm} Messiah A. M. L.: Quantum Mechanics. North-Holland, Amsterdam (1960)

\bibitem{dirac} Dirac P. A. M.: Principles of Quantum Theory. Dover, New York (1970)

\bibitem{schlipp} Schlipp P. A. (Ed.): Albert Einstein: Philosopher Scientist, Open Court Publishing Company
- The Library of Living Philosophers, Vol VII, La Salle. (1970)

\end{thebibliography}
\end{document}